\documentclass[3p,times,procedia]{elsarticle}
\usepackage{nupha_ecrc}

\volume{00}

\firstpage{1}

\journalname{Nuclear Physics A}

\runauth{K. Dusling, M. Mace, V. V. Skokov, P. Tribedy, R. Venugopalan}

\jid{nupha}

\jnltitlelogo{Nuclear Physics A}

\usepackage{amssymb}
\usepackage{subcaption}
\usepackage{graphicx}

\usepackage[figuresright]{rotating}

\begin{document}

\begin{frontmatter}



\dochead{XXVIIth International Conference on Ultrarelativistic Nucleus-Nucleus Collisions\\ (Quark Matter 2018)}

\title{Multiparticle correlations and collectivity in small systems from the initial state}


\author[APS,BNL]{Kevin Dusling}
\author[BNL,SBU]{Mark Mace}
\author[RBRC]{Vladimir V. Skokov}
\author[BNL]{\\Prithwish Tribedy}
\author[BNL]{Raju Venugopalan}

\address[APS]{American Physical Society, 1 Research Road, Ridge, NY 11961, USA}
\address[BNL]{Physics Department, Brookhaven National Laboratory, Bldg. 510A, Upton, NY 11973, USA}
\address[SBU]{Physics and Astronomy Department, Stony Brook University, Stony Brook, NY 11974, USA}
\address[RBRC]{RIKEN-BNL Research Center, Brookhaven National Laboratory, Bldg. 510A, Upton, NY 11973, USA}

\begin{abstract}
We report on recent progress in understanding multiparticle correlations in small systems from the initial state. First, we consider a proof-of-principle parton model, which we use to demonstrate that many of the multiparticle correlations observed in light-heavy ion collisions, often ascribed to hydrodynamic collectivity, can be qualitatively reproduced in an initial state model. Then, we study the two-particle harmonics $v_2$ and $v_3$ for p/d/$^3$He+Au collisions at RHIC using the dilute-dense Color Glass Condensate Effective Field Theory framework. We show that this provides a viable alternative explanation to hydrodynamics, and elaborate on how such modeling can be improved.
\end{abstract}




\end{frontmatter}


\section{Introduction}
\label{sec:intro}
Multiparticle correlations observed in larger colliding nuclear systems such as gold--gold at RHIC and lead--lead at the LHC have, in part, led to the discovery of nearly perfect fluid, the Quark Gluon Plasma (QGP). These correlations are characterized as $n$-th azimuthal moments of the $m$-particle inclusive distribution, the $v_n\{m\}$ harmonics. For larger systems, the observation of QGP formation leads to the paradigm that the system can be accurately modeled as a relativistic viscous fluid. While the dominance of final state effects, efficiently encoded into a hydrodynamic description, limits the influence of correlations from the initial state and the earliest times, smaller systems, like proton--lead collisions at the LHC or proton/deuteron/helium-3--gold collisions at RHIC, offer a unique window to study such the physics of the initial state. These smaller systems strikingly exhibit similar correlations to that of the larger systems~\cite{Aaboud:2017acw}. However, the apparent lack of jet quenching in small systems~\cite{Adam:2016dau}, a strong signal for the existence of a QGP in larger systems, begs the question if it is possible that these prima facie similarities could originate from formally distinct mechanisms in small and large systems. In this proceeding, we will report on recent work in this direction within the Color Glass Condensate (CGC) Effective Field Theory (EFT) framework~\cite{Gelis:2010nm}. The CGC EFT describes nuclei at high energies as composed primarily of high density gluons, whereby an emergent semi-hard scale in each nucleus forms, $Q_s$, which becomes the dominant scale of the problem. Gluons with $k_\perp<Q_s$ are highly occupied, forming strongly correlated domains of size $1/Q_s$. This picture however receives corrections in the full CGC EFT framework as we will see.

This proceeding is organized as follows: in Sec.~\ref{sec:parton} we use a proof-of-principle parton model for multiparticle correlations~\cite{Dusling:2017dqg,Dusling:2017aot,Dusling:2018hsg} to study pA collisions. We show that the similarity of the multiparticle $v_2\{m\geq4\}$ is not unique to a hydrodynamic description. In Sec.~\ref{sec:ddgluons}, we detail an event-by-event model for gluon correlations using the dilute-dense CGC framework~\cite{Mace:2018vwq} to study the ``geometry scan'' recently reported by the PHENIX collaboration at RHIC~\cite{Aidala:2018mcw}, suggesting that the CGC EFT framework can produce ``flow'' in line with observation.

\section{Parton model and collectivity}
\label{sec:parton}
Following~\cite{Lappi:2015vha,Lappi:2015vta}, we consider a model of nearly collinear ($k_\perp\simeq0$) quarks eikonal scattering off of a gluon dense nuclear target within the dilute-dense approximation\footnote{Formally this limit refers to the power-counting in target and projectile color charge densities~\cite{Dumitru:2001ux}, where all orders in the target are considered but only leading order in the projectile, however intuitively it can be understood as $Q_s^2({\rm target}) \gg Q_s^2({\rm projectile})$.}~\cite{Dusling:2017dqg,Dusling:2017aot,Dusling:2018hsg}. We consider only the saturation momentum of the target, $Q_{s,T}=Q_s$. The scattering amplitude of an eikonal quark at transverse position $\mathbf{x}_\perp$ coherently multiple scattering off of the target classical color field is represented by the Wilson line $U(\mathbf{x}_\perp)$.
For simplicity, we assume the Wigner function for quarks in the projectile to be Gaussian in both position and transverse momentum, and described by a single fixed scale $B_p=4~{\rm GeV}^{-2}$~\cite{Kowalski:2006hc}. With these building blocks, we can define an $m$-quark inclusive gluon distribution, $\left\langle\frac{d^{m} N}{d^2\mathbf{p_{1}}\cdots d^2\mathbf{p_{m}} } \right\rangle$ (the full expression and derivation is provided in~\cite{Dusling:2017aot}).
Expectation values are taken over Gaussian color configurations within the McLerran-Venugopalan model~\cite{McLerran:1993ka,McLerran:1993ni}, and over all events. 
Calculating the $p_\perp$-integrated intrinsic $m$-particle $n$-th azimuthal correlations, $\left< e^{i n(\phi_1+\cdots+\phi_{m/2-1}-\phi_{m/2}-\cdots\phi_{m})} \right>$, using the $m$-particle inclusive distribution,  
allows us to compute cumulants $c_n\{m\}$ and the corresponding Fourier harmonics $v_n\{m\}$ in the standard way~\cite{Borghini:2001vi}. Our model has three free parameters: $Q_{s,T}$, $B_p$, and the maximum integrated quark momentum $p_\perp^{\rm max}$\footnote{The value of $p_\perp$ for the final state quark in our parton model is dominantly from the transverse momentum kick  from the target.}.

Remarkably, this simple model is able to capture many of the features observed experimentally. Key among these was the realization of a finite $v_2\{4\}$ for the first time from a systematic initial state calculation. Comparing to the single scattering limit, the Glasma graph approximation~\cite{Dumitru:2008wn} (see e.g.~\cite{Dusling:2013qoz} and references therein), we find that coherent multiple scattering is the salient feature for a finite $v_2\{4\}$~\cite{Dusling:2017aot}.

\begin{figure}
\centering
\begin{subfigure}[t]{0.45\textwidth}
\centering
\includegraphics[width=\textwidth]{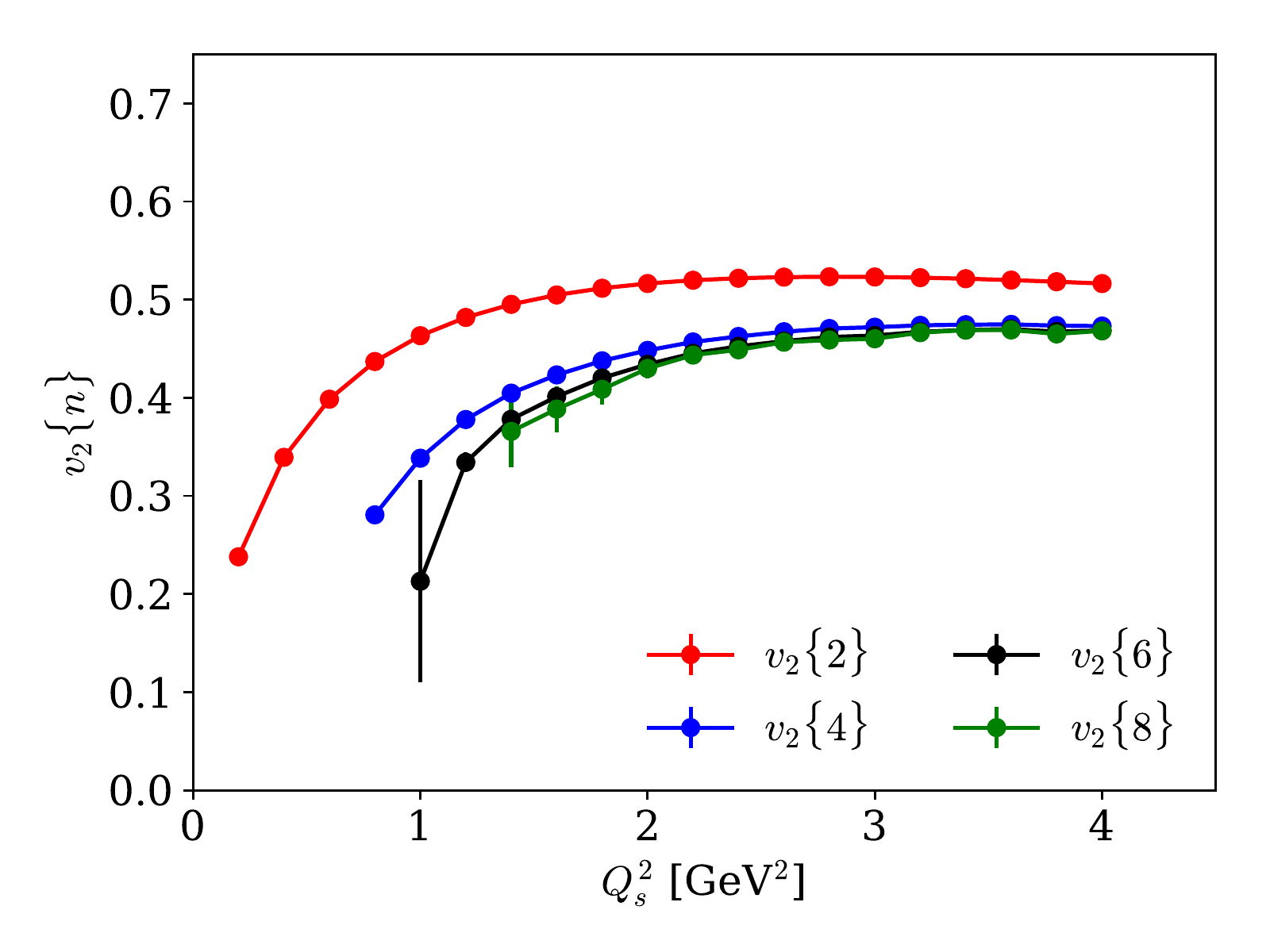}
\label{fig:abelian}
\end{subfigure}
\begin{subfigure}[t]{0.45\textwidth}
\centering
\includegraphics[width=\textwidth]{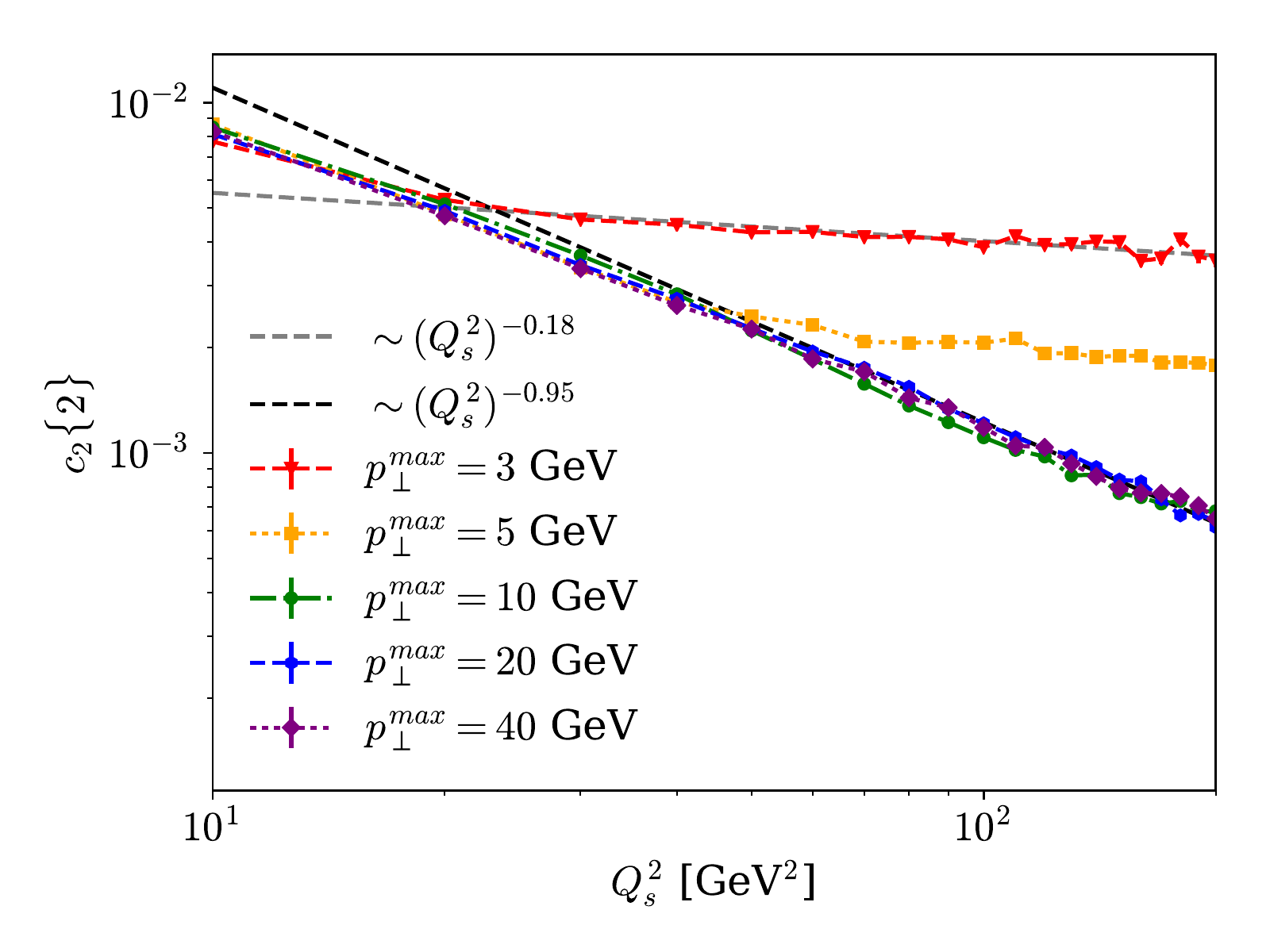}
\label{fig:c22_domains}
\end{subfigure}
\caption{Left: Parton model calculation for coherent multiple scattering off an Abelian field of two, four, six, and eight particle Fourier harmonics as a function of $Q_{s,T}^2$ for $p_\perp^{\rm max}=3~\rm{GeV}$. Right: Probe resolution scale dependence; see text for discussion.}
\label{fig:abelandc2}
\end{figure}

However, calculations of greater than four-particle correlations, even within this simple model, are computationally prohibitive. To proceed, we consider a further simplification, where quarks coherently multiple scatter off an Abelian field\footnote{This reduces traces of products of SU(3) matrices to complex scalar multiplication~\cite{Dusling:2017aot}.}. Results for up to eight particle correlations are shown in left panel of Fig.~\ref{fig:abelandc2}. We plot as a function of $Q_{s,T}^2$, which directly depends on center-of-mass beam energy. The multiplicity however only logarithmically depends on $Q_{s,T}$~\cite{Dumitru:2001ux}; in order to study the multiplicity dependence, more realistic modeling is required, as we describe in the next section. Nevertheless, this clearly demonstrates that the similarity of $v_2\{m\geq4\}$, commonly believed to be a signature for the formation of a hydrodynamic medium, can also be generated from the initial state. 

By increasing the value of $Q_{s,T}^2$, na\"{i}vely this increases the number of domains, $N_D\sim B_p Q_{s,T}^2$, in the target seen by the projectile. However, the correlations in left panel of Fig.~\ref{fig:abelandc2} do not fall off as $1/N_D$ as would be the case for an truly independent domain model. To elucidate why this is, we consider the full non-Abelian version of our model the in right panel of Fig.~\ref{fig:abelandc2}, where we vary the probe resolution scale, given by the dimensionless quantity $Q_{s,T}^2/(p_\perp^{\rm max})^2$. For $(p_\perp^{\rm max})^2 \ll Q_{s,T}^2$, the probe coarse-grains over multiple domains, and thus the falloff is much slower than $1/N_D$. For $(p_\perp^{\rm max})^2 \gg Q_{s,T}^2$, the probe is able to resolve areas less than the domain size, and thus sees each domain individually, resulting in a fall off $\sim1/N_D$.

Major shortcomings of this model are considering quarks instead of gluons, which are known to dominate the nuclear wavefunction at high energies, as well as the simplicity of the nuclei modeling -- we will explore both of these issues in the next section.

\section{Dilute-dense gluons and PHENIX system scan}
\label{sec:ddgluons}
We now consider a more physical model of correlations of gluons in the dilute-dense limit of the CGC EFT~\cite{Dumitru:2001ux}. This is possible due to the recent observation that the first non-trivial correction in the color charge density of the projectile produces finite odd harmonics~\cite{McLerran:2016snu,Kovchegov:2018jun}. 
We consider the recent system scan of p/d/$^3$He+Au by the PHENIX collaboration~\cite{Aidala:2018mcw}. They posit that the observed ordering for $v_2$ and $v_3$ is a clear signature of collectivity driven by the projectile geometry, the initial ellipticity or triangularity, which is captured by hydrodynamic response. However we will show this is not the only viable explanation.

First, the initial color charge densities are determined analogously to the IP-Glasma model~\cite{Schenke:2012wb}.
Given the two color charge configurations, we then calculate the single inclusive distribution, for details see~\cite{Mace:2018vwq}. Taking configuration and event averages allows us to define the multiplicity as well as multiparticle distributions. The calculated multiplicity distribution is a convolution of negative binomial distributions -- this is not an input. All free parameters, listed in detail in~\cite{Mace:2018vwq}, are constrained by minimizing deviations from the STAR d+Au multiplicity distribution~\cite{Abelev:2008ab}. We find that good agreement up to very large multiplicities~\cite{Mace:2018vwq}.

\begin{figure}
\centering
 \includegraphics[width=\textwidth]{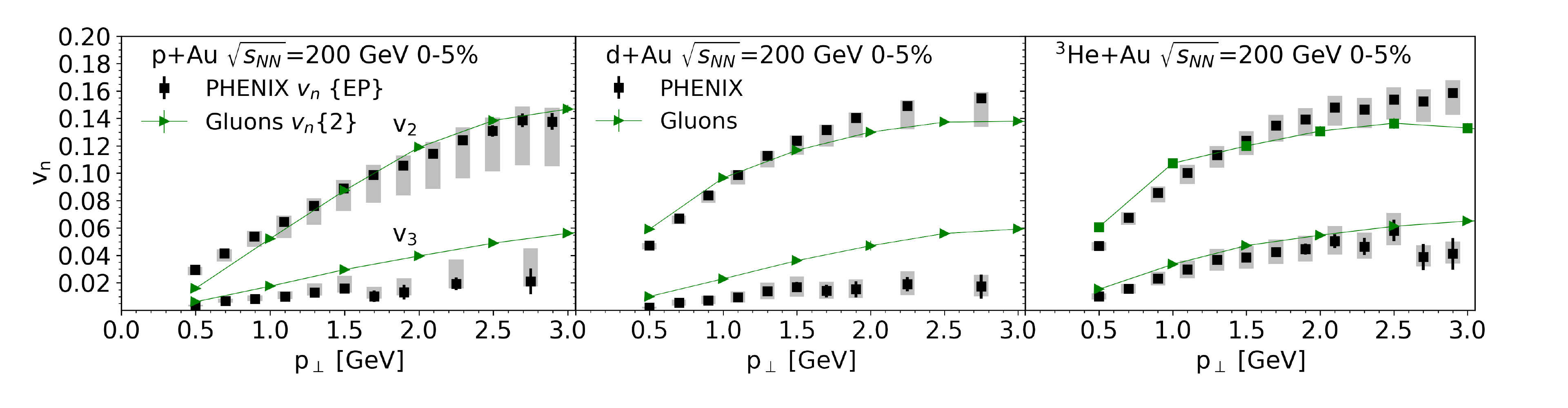}
\caption{Dilute-dense gluon correlations for $v_{2}$ and $v_{3}$ compared to recent PHENIX measurements~\cite{Aidala:2018mcw}.}
\label{fig:ddphenix}
\end{figure}

We then calculate the two-particle $v_{2,3}$ for p/d/$^3$He+Au collision at RHIC energies. The results are shown in Fig.~\ref{fig:ddphenix}. For $v_2$, the dilute-dense calculation describes the data well, while for $v_3$ p+Au and d+Au are overestimated.  In our framework, the ordering comes from the fact that for the 0-5\% centrality class, the different systems probe different multiplicities, and thus different integrated $Q_s$ are probed, which controls the strength of the correlations. Our results, consistent with experiment, are opposite to na\"{i}ve domain expectations due to the effects of coherence and domain resolution, previously discussed in the explanation of the right panel of Fig.~\ref{fig:abelandc2}. The plotted uncertainties are only statistical; systematic uncertainties remain to be quantified. We must understand the impact of dense projectile effects, which can be studied by comparison to classical Yang-Mills dynamics calculations and will be reported elsewhere~\cite{Bjoern-Chun-Mark-Prithwish-Vladi-Raju}. Additionally, uncertainties exist from system parameter dependence -- we fix all parameters for all systems using the STAR d+Au multiplicity distribution. Having p+Au and $^3$He+Au multiplicity distributions would eliminate this uncertainty, and is also important for hydrodynamic modeling where they are used as inputs. Lastly, we considered the correlations of gluons, and not fragmented hadrons. It is necessary to include fragmentation in the future; a promising approach was recently developed in~\cite{Schenke:2016lrs}.
\vspace{-0.5cm}
\section{Conclusions}

While on the surface small systems appear very similar to larger high-energy nuclear collisions, key features of a strongly interacting plasma like jet quenching remain elusive, encouraging further study. We have demonstrated that many of the observed multiparticle correlations can be qualitatively, and perhaps even quantitatively, reproduced by purely initial state models. While quantitative initial state modeling of small systems is still in its infancy due to the complexity and theoretical uncertainties, this is a promising avenue of study. Recently, we have shown that the framework of Sec.~\ref{sec:ddgluons} can also describe the multiplicity dependence of $v_{2,3,4}$ at the LHC~\cite{Mace:2018yvl}. Looking forward, we plan to quantify the differences between initial state dilute/dense-dense frameworks and the relative contribution of a possible short-lived fluid stage~\cite{Bjoern-Chun-Mark-Prithwish-Vladi-Raju}.
{\it Acknowledgements:} We thank Sylvia Morrow for providing us the PHENIX data presented in Fig.~\ref{fig:ddphenix}. This material is based on work supported by the U.S. Department of
Energy, Office of Science, Office of Nuclear Physics, under Contracts No.
DE-SC0012704 (M.M.,P.T.,R.V.) and DE-FG02-88ER40388 (M.M.). M.M. would also
like to thank the BEST Collaboration for support. This research used resources
of the National Energy Research Scientific Computing Center, a DOE Office of
Science User Facility supported by the Office of Science of the U.S. Department
of Energy under Contract No. DE-AC02-05CH11231.





\bibliographystyle{elsarticle-num}
\bibliography{bibl}







\end{document}